\documentclass[preprint,12pt]{elsarticle}

\usepackage{amssymb}
\usepackage{color}

\newtheorem{theorem}{Theorem}

\newtheorem{remark}{Remark}
\newtheorem{lemma}{Lemma}
\newtheorem{assumption}{Assumption}

\journal{\ \ \ }

\begin{document}

\begin{frontmatter}



\title{Distributed Second-order Multi-Agent Optimization over unbalanced network without boundedness of gradients\tnoteref{mytitlenote}}
\tnotetext[mytitlenote]{This work is supported by National Natural Science
Foundation (NNSF) of China (Grant Nos. 61973329 and 61772063) and the Beijing Natural Science Foundation (Grant Nos. Z180005 and 9192008).}


\author{Lipo Mo}
\ead{beihangmlp$@$126.com}
\author{Haokun Hu}
\address{School of Mathematics and Statistics,
        Beijing Technology and Business University, Beijing 100048, P.~R.~China}
\author{Yongguang Yu}
\author{Guojian Ren}
\address{Department of Mathematics,
        Beijing Jiaotong University, Beijing 10044, P.~R.~China}

\begin{abstract}
This paper is mainly devoted to the distributed second-order multi-agent optimization problem with unbalanced and directed networks. To deal with this problem, a new distributed algorithm is proposed based on the local neighbor information and the private objective functions. By a coordination transformation, the closed-loop system is divided into two first-order subsystems, which is easier to be dealt with. Under the assumption of the strong connectivity of networks, it is proved that all agent can collaboratively converge to some optimal solution of the team objective function, where the gradient of the private objective functions is not assumed to be bounded. 
\end{abstract}


\end{frontmatter}


\section{Introduction}

Recently, the distributed optimization (see  \cite{survey1,survey2}) has drawn an increasing interest due to its extensive applications, such as machine learning \cite{learning}, distributed regression \cite{regression} and distributed resource allocation \cite{resource}. Based on the multi-agent networks, many distributed algorithms were proposed to drive agents collaboratively seek the optimal solution of an objective function, which is the sum of local objective functions. For example, a distributed subgradient algorithm was introduced in \cite{Nedic2009} to tackle the unconstrained optimization problem when the directed communication network was balanced and strongly connected. Then, this algorithm was extended to the situations of identical or nonidentical constrains by employing the projection operation in \cite{Nedic2010,Lin2016}, especially, in \cite{Lin2016}, it was also proved that the distributed optimization could not be achieved by this algorithm when the communication network was not balanced. To improve the convergence rate of the proposed algorithm, fixed gradient gains, fixed step sizes and sign function were introduced to the algorithms in \cite{Liuqingshan,Lin2017optimal,Leijinlong} and it was proved that the distributed optimization problem could be solved under the assumption of undirected and connected networks. When the constraint sets were unbounded, two distributed algorithms were designed in \cite{Lin2019optimal2} to solve the continuous-time and discrete-time multi-agent optimization problem. For second-order multi-agent networks, the authors of \cite{Zhang2014,Liu2015,Mo2019} proposed some distributed algorithms with uniform or nonuniform gradient gains to deal with the distributed optimization problem when the communication network was undirected and connected. Besides, some classical algorithms were modified to solve the situation of directed and balanced or detailed balanced networks in \cite{Wangdong,Wangaijuan}, where the objective function was assumed to be strongly convex. When the uncertainty of parameters and nonconvex input constraint were taken into account, some new algorithms with adaptive term and nonconvex constraint were designed to deal with these situations in \cite{MoCao2019,MoZhao2019,Lin2019optimal1}, where the objective functions were not assumed to be strongly convex.

All aforementioned works were assumed that the communication network was undirected or balanced, which is hard to be implemented in real plant due to the complexity and uncertainty of environment.  When the directed communication graph was unbalanced, the authors of \cite{Nedic2015} combined the subgradient algorithm and push-sum mechanism to deal with the time-varying network optimization problem and also gave the convergence rate of the proposed algorithm. Inspired by surplus consensus algorithm \cite{Cai2012}, the authors of \cite{Xi2017,Xi2018,Mai2019,WangLei2019} designed some distributed protocols to force all agents achieve distributed optimization under the assumption that the gradient (subgradient) was bounded or the local objective function was strongly convex. Then, these algorithms were modified in \cite{Xiepei2018,Lihuaqing2019} based on epigraph and projection operator to solve the time-varying network optimization problem with convex constraints and inequality constraints. In \cite{Liang20191,Liang20192}, quasi-monotone algorithm and dual averaging push algorithm were introduced to seek the optimal solution of the nonsmooth objective function. Note that most of these works were assumed that the dynamics of all agents were first-order integrator and the objective function was strongly convex or the gradient (subgradient) was bounded. In real applications, however, the dynamics of agents might not be modeled by first-order integrator because it is more convenient to control the force (accelerate) of agent \cite{MoLin2018,Caoer,Xiang}, which can be modeled by second-order dynamics. It is more challengeable to study the distributed second-order multi-agent optimization due to the complexity of dynamics. Besides, the assumption of boundedness of the gradient is too strict, and many objective function doesn't satisfy this assumption, especially unconstrained optimization problem.

This paper primarily investigates the distributed second-order multi-agent optimization problem with unbalanced and directed networks. A novel distributed algorithm inspired by surplus consensus scheme is designed. Then we prove that all agents can collaboratively find one optimal solution of the team objective function. The main contribution of this paper is composed of the following three aspects.

First, compared with \cite{Xi2017,Xi2018,WangLei2019,Liang20191,Liang20192}, where the distributed first-order multi-agent optimization were considered, this paper considers the second-order optimization problem. To conquer the difficulties from the higher dynamics, a new coordination transformation is introduced to change the closed-loop system into two parts, which is easier to be analyzed.

Second, compared with \cite{Lin2016,Liuqingshan,Lin2017optimal,Lin2019optimal2,MoCao2019,MoZhao2019,Lin2019optimal1}, where some distributed algorithms were proposed to solve the optimal problem over undirected or balanced networks, this paper deals with the situation of unbalanced and directed network. Due to the lack of double stochasticity of system matrix, many classical analysis methods do not work here, which brings us more challenge when we prove the convergence of the proposed algorithm.

Third, compared with \cite{Xi2017,WangLei2019}, where it was assumed that the gradient (subgradient) was bounded, this paper removes this restriction, which enlarges the applicable range of the proposed algorithm. Besides, the proof methods based on Lyapunov functions for boundedness in \cite{Lin2019optimal2,Lin2019optimal1} are invalid due to the linear iteration doesn't satisfy convexity, we introduce a new method to show the boundedness of the system states and gradients.

\section{Preliminaries and Problem Statement}
Let $\mathcal{G} = (V, E)$ be a digraph, where $V = \{1, 2, \cdots, n\}$ is the node set and $E$ is the directed edge set. The adjacent matrix $A = [a_{ij}]$ is a nonnegative matrix, where $a_{ij} > 0$ if and only if $(j, i) \in E$. The neighbor set of node $i$ is defined as $N_i = \{j \in V | a_{ij} > 0\}$ and the out-neighbor set is defined as $N_i^{out} = \{j \in V | a_{ji} > 0\}$. The Laplacian $L$ of $G$ is a square matrix, where $[L]_{ij} = - a_{ij}$ for $i \neq j$ and $[L]_{ii} = \sum\limits_{j=1}^{n} a_{ij}$. We say there is a directed path from node $i$ to node $j$ if there exist $i_1, i_2, \cdots, i_k \in V$ such that $(i, i_1), (i_1, i_2), \cdots, (i_{k-1}, i_k), (i_k, j) \in E$. The digraph $G$ is balanced if $\sum\limits_{j=1}^n a_{ij} = \sum\limits_{i=1}^n a_{ij}$, otherwise, we say the digraph $G$ is unbalanced. One can see \cite{graph} for details about graph theory.

Consider the following second-order multi-agent network with $n$ agents. Suppose the dynamical equation of each agent is:
\begin{equation}
\begin{array}{lll}
r_i(k+1)=r_i(k)+q_i(k)T,\\[0.2cm]
q_{i}(k+1)=q_i(k)+u_i(k)T, \ \ i\in V, \label{unbalance1}
\end{array}
\end{equation}
where $r_i(k), q_i(k), u_i(k)\in \mathbf{R}^s$ are the position state, velocity state and control input of the $i$th agent at time $k$, $T > 0$ is the sampling time.

The brief aim of this paper is to design a control algorithm  to drive all agents collaboratively find one optimal point of the following team objective function, which is the sum of some private objective functions:
\begin{equation}\label{mubiao}
\min\limits_{s \in \mathbf{R}^{s}} f(s) = \sum\limits_{i=1}^{n} f_i(s),
\end{equation}
where $f_i(s): \mathbf{R}^s \rightarrow \mathbf{R}$ is both differentiable and convex, the $i$th agent can only receive the information of $f_i(s)$ and can not receive the information of $f(s)$. Let $X$ be the optimal solution set of $f(s)$.

\section{Algorithm Design}

Define $b_{ij} > 0$ if $i \in N_j^{out}$, otherwise $b_{ij} = 0$ for $i \neq j$. Set $b_{ii} = 1 - \sum\limits_{j \in N_i^{out}} b_{ij} > 0$. Clearly, $\sum\limits_{i=1}^n b_{ij} = 1$. In this paper, we propose the following algorithm:
\begin{equation}\label{algorithm}
\begin{array}{lll}
u_i(k) = - q_i(k) + \sum\limits_{j\in N_i} a_{ij} [(r_j(k) - r_i(k)) + (q_j(k) - q_i(k))] + \epsilon y_i(k)\\[0.2cm]
y_i(k+1) = - \sum\limits_{j\in N_i} a_{ij} [(r_j(k) - r_i(k)) + (q_j(k) - q_i(k))]T + \sum\limits_{j=1}^n b_{ij} y_j(k)\\[0.2cm]
\ \ \ \ \ \ \ \ \ \ \ \ \ \  - \epsilon y_i(k) T - \alpha_k d_i(k) T,\\[0.2cm]
d_i(k) = \left\{
           \begin{array}{ll}
             0, & \frac{1}{\sqrt{\alpha_k}} < \|\nabla f_i(w_i(k))\|^2 \\
             \nabla f_i(w_i(k)), &  {otherwise},
           \end{array}
         \right.\\[0.2cm]
w_i(k) = r_i(k) + q_i(k) + \sum_{j=1}^n b_{ij} y_j(k),
\end{array}
\end{equation}
where $y_i(k), d_i(k)$ and $w_i(k)$ are the auxiliary variables with appropriate dimension, $\epsilon > 0$ is sufficient small \cite{Xi2017} and $\alpha_k > 0$ is the stepsize of gradient.
Let $x_i(k) = r_i(k) + q_i(k)$. Then
\[
\begin{array}{lll}
r_i(k+1) = r_i(k) + (x_i(k) - r_i(k))T = (1-T) r_i(k) + T x_i(k)\\[0.2cm]
x_i(k+1) = x_i(k) + \sum\limits_{j\in N_i} a_{ij}T (x_j(k) - x_i(k)) + \epsilon y_i(k) T\\[0.2cm]
y_i(k+1) = -\sum\limits_{j\in N_i} a_{ij}T (x_j(k) - x_i(k)) - \epsilon y_i(k) T + \sum\limits_{j=1}^n b_{ij} y_j(k)-\alpha_k d_i(k)T.
\end{array}
\]
Define $z_i(k) = x_i(k)$ and $z_{n+i}(k) = y_i(k)$ for all $i = 1, 2, \cdots, n$, $Z(k) = [z_1(k)^T, \cdots, z_{2n}(k)^T]^T$, $M = \left[
                                                                  \begin{array}{cc}
                                                                    I - LT & \epsilon T I \\
                                                                    LT & B - \epsilon T I \\
                                                                  \end{array}
                                                                \right] \in \mathbf{R}^{2n\times 2n}
$, $\nabla F(k) = [0^T, \cdots, 0^T, d_1(k)^T, \cdots, d_n(k)^T]^T$. Then the closed-loop system can be rewritten as
\begin{equation}\label{unbalance5}
r_i(k+1) = (1 - T) r_i(k) + T x_i(k),
\end{equation}
\begin{equation}\label{unbalance6}
Z(k+1) = [M\otimes I_s] Z(k) - \alpha_k T \nabla F(k).
\end{equation}

To proceed our analysis, we need the following assumptions.
\begin{assumption}\label{as1}\rm
Suppose $X_i \neq \emptyset$ is bounded for all $i = 1, 2, \cdots, n$, where $X_i = \{s | \nabla f_i(s) = 0\}$.
\end{assumption}
\begin{remark}\rm
Since $f_i(s)$ is both differentiable and convex, $X_i$ is the optimal set of $f_i(s)$. By Lemma 1 in \cite{Lin2017optimal}, $X \neq \emptyset$ and $X$ is a closed convex set. Under Assumption \ref{as1}, it is easy to prove that $\lim\limits_{\|x\| \rightarrow \infty} f_i(x) = +\infty$.
\end{remark}
\begin{assumption}\label{as2}\rm
The communication graph is strongly connected.
\end{assumption}
\begin{remark}\rm
In this paper, we only require that the communication graph $G$ is strongly connected, not necessary to be balanced. While, in \cite{Lin2019optimal2,Mo2019,MoZhao2019}, it is assumed that the communication graph is also balanced. In the future work, we will study the situation of switching graphs\cite{Caoqie}.
\end{remark}
\begin{assumption}\label{as4}\rm
Suppose the gradient gains $\{\alpha_k\}$ satisfies $\alpha_k \geq \alpha_{k+1}$ for all $k$, $\sum\limits_{k=1}^\infty \alpha_k = \infty$ and $\sum\limits_{k=1}^\infty \alpha_k^2 < \infty$.
\end{assumption}
\begin{remark}\rm
Assumption \ref{as4} implies that $\lim\limits_{k\rightarrow\infty} \alpha_k = 0$, i.e., the effect of the gradient is diminishing. The condition $\sum\limits_{k=1}^\infty \alpha_k = \infty$ means the gradient term works persistently.
\end{remark}

The following lemma establishes the convergence rate of $M^k$, which is useful to analyze the convergence of the proposed algorithm.
\begin{lemma}\rm\cite{Xi2017}\label{lem1}
Under Assumption \ref{as2},
there exist $\Gamma > 0$ and $0 < \gamma < 1$, such that $\|M^k - \left[
                                             \begin{array}{cc}
                                               \frac{1}{n}\mathbf{1}_n\mathbf{1}_n^T & \frac{1}{n}\mathbf{1}_n\mathbf{1}_n^T \\
                                               0 & 0 \\
                                             \end{array}
                                           \right]\| \leq \Gamma \gamma^k$.
\end{lemma}

\section{Convergence Analysis}

In this section, we mainly prove the convergence of the proposed algorithm. First, let us prove the boundedness of the states of system (\ref{unbalance6}).
\begin{lemma}\rm\label{lem2}
Suppose Assumptions  \ref{as1}, \ref{as2} and \ref{as4} hold. Then the states of system (\ref{unbalance6}) are bounded if $\frac{1}{T} - \sum_{j\in N_i} a_{ij} > 0$ and $\epsilon < \frac{1}{T} - \sum_{j\in N_i} a_{ij}$ for all $i \in V$.
\end{lemma}
{\bf Proof.}
Let $\theta_i(k) = x_i(k) + y_i(k), i \in V$. Then
\[
\begin{array}{lll}
x_i(k+1) = x_i(k) + \sum_{j\in N_i} a_{ij}T(x_j(k) - x_i(k)) + \epsilon T (\theta_i(k) - x_i(k))\\[0.2cm]
\ \ \ \ =(1 - \sum_{j\in N_i} a_{ij}T - \epsilon T) x_i(k) + \sum_{j\in N_i} a_{ij}T x_j(k) + \epsilon T \theta_i(k), \\[0.2cm]
\theta_i(k+1) = x_i(k) + \sum_{j=1}^n b_{ij} [\theta_j(k) - x_j(k)] - \alpha_k T d_i(k)\\[0.2cm]
\ \ \ \ =w_i(k) - \alpha_k T d_i(k).
\end{array}
\]
Define $V_1(k) = \max_i\{\|x_i(k) - s\|^2, \|\theta_i(k) - s\|^2\}$, where $s \in X$. By the convexity of the square function, we have
\[
\begin{array}{lll}
\|x_i(k+1) - s\|^2 = \|(1 - \sum_{j\in N_i} a_{ij}T - \epsilon T) [x_i(k)-s]\\[0.2cm]
\ \ \ \ \ \ \ \  + \sum_{j\in N_i} a_{ij}T [x_j(k)-s] + \epsilon T [\theta_i(k)-s]\|^2\\[0.2cm]
\leq (1 - \sum_{j\in N_i} a_{ij}T - \epsilon T) \|x_i(k)-s\|^2 + \sum_{j\in N_i} a_{ij}T \|x_j(k)-s\|^2\\[0.2cm]
\ \ \ \ \ \ \ \  + \epsilon T \|\theta_i(k)-s\|^2\\[0.2cm]
\leq V_1(k).
\end{array}
\]
If $d_i(k) = 0$, then,
\[
\begin{array}{lll}
\|\theta_i(k+1) - s\|^2 = \|(1-b_{ii}) [x_i(k) - s] + \sum_{j=1}^n b_{ij} [\theta_j(k) - s]\\[0.2cm]
\ \ \ \ \ \ \ \ \ \ \ \ \ \  + \sum_{j\neq i}b_{ij} [x_j(k) - s]\|^2\\[0.2cm]
\leq (1-b_{ii}) \|x_i(k) - s\|^2 + \sum_{j=1}^n b_{ij} \|\theta_j(k) - s\|^2 + \sum_{j\neq i}b_{ij} \|x_j(k) - s\|^2\\[0.2cm]
\leq V_1(k).
\end{array}
\]
If $d_i(k) \neq 0$, then $\|\nabla f_i(w_i(k))\|^2 \leq \alpha_k^{-\frac{1}{2}}$. Thus
\[
\begin{array}{lll}
&&\|\theta_i(k+1) - x\|^2\\[0.2cm]
 &=& \|w_i(k) - x - \alpha_k T d_i(k)\|^2\\[0.2cm]
&=& \|w_i(k) - x\|^2 + \alpha_k^2 T^2 \|d_i(k)\|^2 - 2(w_i(k) - x)^T \alpha_k T \nabla f_i(w_i(k))\\[0.2cm]
&\leq& \|w_i(k) - x\|^2 + \alpha_k^2 T^2 \|d_i(k)\|^2 - 2 \alpha_k T [f_i(w_i(k))-f_i(x)]\\[0.2cm]
&=& \|w_i(k) - x\|^2 + \alpha_k^2 T^2 \|\nabla f_i(w_i(k))\|^2\\[0.2cm]
&& \ \ \ \ \ \ \ \ \ \ \ \  - 2 \alpha_k T [f_i(w_i(k))-f_i(x_i) + f_i(x_i)-f_i(x)]\\[0.2cm]
&\leq& \|w_i(k) - x\|^2 + \alpha_k^{\frac{3}{2}} T^2 + 2 \alpha_k T [f_i(x)-f_i(x_i)] \\[0.2cm]
&& \ \ \ \ \ \ \ \ \ \ \ \ - 2 \alpha_k T [f_i(w_i(k))-f_i(x_i)].
\end{array}
\]
Since $X$ and $X_i$ are all nonempty and bounded, we can choose $L > 0$ large enough such that $X \subset Y_1$, $X_i \subset Y_1$, $w_i(0) \in Y_1$ and $f_i(w_i(k)) - f_i(x_i) \geq \frac{1}{2} \alpha_k^{\frac{1}{2}} T + f_i(x) - f_i(x_i)$ for all $w_i(k) \notin Y_1$, $x \in X$ and $x_i \in X_i$ for all $i = 1, 2, \cdots, n$, where $Y_1 = \{\xi \in \mathbf{R}^s | \|\xi - x\| \leq L\}$. If $w_i(k) \in Y_1$, then $\|\theta_i(k+1) - x\|^2 \leq L + T^2 \alpha_k^{\frac{3}{2}} + 2 \alpha_k T [f_i(x) - f_i(x_i)]$. Since $\lim_{k\rightarrow\infty} \alpha_k = 0$ and $f_i(x)$ and $f_i(x)$ are bounded because $X$ and $X_i$ are closed and bounded, $L + T^2 \alpha_k^{\frac{3}{2}} + 2 \alpha_k T [f_i(x) - f_i(x_i)]$ is bounded. Hence, $\|\theta_i(k + 1) - x\|$ is bounded. If $w_i(k) \notin Y_1$, then $\|\theta_i(k + 1) - x\|^2 \leq \|w_i(k) - x\|^2 \leq V_1(k)$. From the above analysis, we can conclude that $V_1(k)$ is bounded. Note that $x \in X$ is bounded, we have $x_i(k)$ and $\theta_i(k)$ are bounded for all $i$. By the definition of $\theta_i(k)$, we can see that $y_i(k)$ is also bounded for all $i \in V$.

\begin{theorem}\rm
Consider the second-order multi-agent network (\ref{unbalance1}). If Assumptions \ref{as1}, \ref{as2} and \ref{as4} hold, then all agents could achieve consensus and the team objective function could also be minimized as time $t$ goes to infinity under Algorithm (\ref{algorithm})  if $\frac{1}{T} - \sum_{j\in N_i} a_{ij} > 0$ and $\epsilon < \frac{1}{T} - \sum_{j\in N_i} a_{ij}$ for all $i \in V$.
\end{theorem}
{\bf Proof.} We first prove that all agents could reach an agreement under Algorithm (\ref{algorithm}). From lemma \ref{lem2}, $x_i(k)$ and $y_i(k)$ are bounded. Hence, $[\nabla F(k)]_i$ is bounded due to the fact that $f_i(x)$ are differentiable for all $i$, i.e., there exists $D > 0$, such that $\|[\nabla F(k)]_i\| \leq D$ for all $i$ and $k$. Note that $\lim\limits_{k\rightarrow\infty} \alpha_k = 0$ and $d_i(k) = \nabla f_i(w_i(k))$ if $\|\nabla f_i(w_i(k))\|^2 \leq \alpha_k^{-\frac{1}{2}}$, there exists $T_0 > 0$ such that $d_i(k) = \nabla f_i(w_i(k))$ for all $k \geq T_0$. In the following, we assume $k \geq T_0$.

For $n + 1 \leq i \leq 2n$, it is clear that $z_i(k) = \sum\limits_{j=1}^{2n} [M^k]_{i,j} z_j(0) - \sum\limits_{r=1}^{k-1}\sum\limits_{j=1}^{2n} [M^{k-r}]_{i,j}$ $\alpha_{r-1} T [\nabla F(r - 1)]_j - \alpha_{r-1} T [\nabla F(k-1)]_i$, where $[ \cdot ]_{ij}$ represents the $(i, j)$ entry of the corresponding matrix and $[ \cdot ]_i$ represents the $i$th block of the corresponding verctor. Since $\lim\limits_{k\rightarrow\infty} \alpha_k = 0$ and $\|\alpha_k [\nabla F(k)]_j\| \leq \alpha_k \|d_i(k)\| \leq \alpha_k^{\frac{3}{4}}$, for any $\epsilon_1 > 0$, there exists $K > T_0$ such that $\|\alpha_k T [\nabla F(k)]_j\| \leq \epsilon_1$ for all $k > K$. By Lemma \ref{lem1}, we have
\[
\begin{array}{lll}
\|z_i(k)\| & \leq & \sum\limits_{j=1}^{2n} \Gamma \gamma^k \| z_j(0) \| + \sum\limits_{r=1}^{K}\sum\limits_{j=1}^{2n} \Gamma \gamma^{k-r} T \alpha_{r-1}^{\frac{3}{4}} +  \sum\limits_{r=K+1}^{k-1}\sum\limits_{j=1}^{2n} \Gamma \gamma^{k-r} \epsilon_1 + \epsilon_1\\[0.2cm]
& \leq & \sum\limits_{j=1}^{2n} \Gamma \gamma^k \| z_j(0) \| + \sum\limits_{r=1}^{K}\sum\limits_{j=1}^{2n} \Gamma \gamma^{k-r} T \alpha_{r-1}^{\frac{3}{4}} +  \frac{2n\Gamma\gamma\epsilon_1}{1-\gamma} + \epsilon_1
\end{array}
\]
for all $k > K +1 $. Hence $\lim\limits_{k\rightarrow\infty} \|z_i(k)\| \leq \frac{2n\Gamma\gamma\epsilon_1}{1-\gamma} + \epsilon_1$. By the arbitrariness of $\epsilon_1$, we can conclude that $\lim\limits_{k\rightarrow\infty} \|z_i(k)\| = 0$ for all $n + 1 \leq i \leq 2n$, which means $\lim_{k\rightarrow\infty} y_i(k) = 0$ for all $1 \leq i \leq n$.

Define $\overline{z}(k) = \frac{1}{n} \sum\limits_{i=1}^{2n} z_i(k) = \frac{1}{n} [\sum\limits_{i=1}^n x_i(k) + \sum\limits_{i=1}^n y_i(k)]$, then $\overline{z}(k+1) = \overline{z}(k) + \frac{1}{n} \sum_{j=1}^{2n} \alpha_k T [\nabla F(k)]_i$.  For $1 \leq i \leq n$, we have
\[
\begin{array}{lll}
z_i(k) = \sum\limits_{j=1}^{2n} [M^k]_{ij} z_j(0) - \sum\limits_{r=1}^{k-1} \sum\limits_{j=1}^{2n} [M^{k-r}]_{ij} \alpha_{r-1} T [\nabla F(r-1)]_j\\[0.2cm]
 \ \ \ \ \ \ \ \ \ \ \ \  - \alpha_{k-1} T [\nabla F(k-1)]_i
\end{array}
\]
and
\[
\begin{array}{lll}
\overline{z}(k) = \sum\limits_{j=1}^{2n} \frac{1}{n} z_j(0) - \sum\limits_{r=1}^{k-1}\sum\limits_{j=1}^{2n} \frac{1}{n} \alpha_{r-1} T [\nabla F(r-1)]_j - \frac{1}{n}\sum\limits_{j=1}^{2n} \alpha_{k-1} T [\nabla F(k-1)]_j.
\end{array}
\]
Hence,
\[
\begin{array}{lll}
\|z_i(k) - \overline{z}(k)\|^2 = [z_i(k) - \overline{z}(k)]^T[z_i(k) - \overline{z}(k)]\\[0.2cm]
=\sum\limits_{j=1}^{2n} ([M^k]_{ij} - \frac{1}{n})z_j(0)^T \sum\limits_{t=1}^{2n} ([M^k]_{it} - \frac{1}{n})z_t(0)\\[0.2cm]
\ \ \ \ +\sum\limits_{r=1}^{k-1}\sum\limits_{j=1}^{2n} ([M^{k-r}]_{ij} - \frac{1}{n})\alpha_{r-1}[\nabla F(r-1)]^T T \sum\limits_{l=1}^{k-1}\sum\limits_{t=1}^{2n} ([M^{k-l}]_{it} - \frac{1}{n})\times\\[0.2cm]
\ \ \ \ \alpha_{r-1}[\nabla F(r-1)]_t T +\alpha_{k-1} [\nabla F(k-1)]_i^T T \alpha_{k-1} [\nabla F(k-1)]_iT\\[0.2cm]
\ \ \ \  + \frac{1}{n^2} \sum\limits_{j=1}^{2n} \alpha_{k-1}[\nabla F(k-1)]_j^T T \sum\limits_{t=1}^{2n} \alpha_{k-1}[\nabla F(k-1)]_tT\\[0.2cm]
\ \ \ \ +2[-\sum\limits_{j=1}^{2n} ([M^{k}]_{ij} - \frac{1}{n}) z_j(0)^T \alpha_{k-1} [\nabla F(k-1)]_iT\\[0.2cm]
\ \ \ \  + \sum\limits_{j=1}^{2n} ([M^{k}]_{ij} - \frac{1}{n}) z_j(0)^T\frac{1}{n} \sum\limits_{t=1}^{2n} \alpha_{k-1} [\nabla F(k-1)]_tT\\[0.2cm]
\ \ \ \ - \sum\limits_{j=1}^{2n} ([M^k]_{ij} - \frac{1}{n}) z_j(0)^T \sum\limits_{r=1}^{k-1} \sum\limits_{t=1}^{2n} ([M^{k-r}]_{it} - \frac{1}{n})\alpha_{r-1} [\nabla F(r-1)]_tT \\[0.2cm]
\ \ \ \ +\sum\limits_{r=1}^{k-1}\sum\limits_{j=1}^{2n} ([M^{k-r}]_{ij} - \frac{1}{n})\alpha_{r-1}[\nabla F(r-1)]_j^T \alpha_{k-1} [\nabla F(k-1)]_iT^2\\[0.2cm]
\ \ \ \ -\sum\limits_{r=1}^{k-1}\sum\limits_{j=1}^{2n} ([M^{k-r}]_{ij} - \frac{1}{n}) \alpha_{r-1} [\nabla F(r-1)]_j^T T^2 \frac{1}{n}\sum\limits_{t=1}^{2n} \alpha_k [\nabla F(k-1)]_t\\[0.2cm]
\ \ \ \  -\alpha_{k-1} [\nabla F(k-1)]_i T \frac{1}{n}\sum\limits_{j=1}^{2n} \alpha_{k-1}[\nabla F(k-1)]_jT]\\[0.2cm]
\end{array}
\]
\[
\begin{array}{lll}
\leq \Gamma^2 \gamma^{2k} (\sum\limits_{j=1}^{2n} \|z_j(0)\|)^2 + 4 \Gamma^2 D^2 n^2 T^2 \sum\limits_{r=1}^{k-1} \alpha_{r-1}\gamma^{k-r}\sum\limits_{l=1}^{k-1}\alpha_{l-1} \gamma^{k-l} + 5D^2 T^2 \alpha_{k-1}^2\\[0.2cm]
\ \ \ \ +2(2n \Gamma^2 D T \gamma^k \sum\limits_{j=1}^{2n} \|z_j(0)\|\sum\limits_{r=1}^{k-1} \gamma^{k-r} \alpha_{r-1} + 3 \Gamma D T \sum\limits_{j=1}^{2n} \|z_j(0)\| \gamma^k \alpha_{k-1}\\[0.2cm]
\ \ \ \ + 4 n \Gamma D^2 T^2 \sum\limits_{r=1}^{k-1} \gamma^{k-r}\alpha_{r-1}\alpha_{k-1} + 2 n \Gamma D^2 T^2 \sum\limits_{r=1}^{k-1} \gamma^{k-r} \alpha_{r-1}\alpha_k + 2 T^2 \alpha_{k-1}^2 D^2).
\end{array}
\]
For any $K' \geq T_0$, $\sum\limits_{k=1}^{K'}\sum\limits_{r=1}^{k-1}\alpha_{r-1}\gamma^{k-r}\sum\limits_{l=1}^{k-1}\alpha_{l-1}\gamma^{k-l} \leq \sum\limits_{k=1}^{K'}\sum\limits_{r=1}^{k-1} \sum\limits_{l=1}^{k-1} \gamma^{k-r} \gamma^{k-l} \frac{1}{2} (\alpha_{r-1}^2 + \alpha_{l-1}^2) = \sum\limits_{k=1}^{K'}\sum\limits_{r=1}^{k-1} \sum\limits_{l=1}^{k-1} \gamma^{k-r} \gamma^{k-l} \alpha_{r-1}^2 \leq \sum\limits_{k=1}^{K'}\sum\limits_{r=1}^{k-1} \gamma^{k-r} \frac{1}{1-\gamma} \alpha_{r-1}^2 \leq \frac{1}{(1-\gamma)^2} \sum\limits_{r=1}^{K'-1} \alpha_{r-1}^2$. Hence,
\[
\begin{array}{lll}
\sum\limits_{k=1}^{K'}\sum\limits_{r=1}^{k-1}\alpha_{r-1}\gamma^{k-r}\sum\limits_{l=1}^{k-1}\alpha_{l-1}\gamma^{k-l} \leq \frac{1}{(1-\gamma)^2} \sum\limits_{k=0}^{K'} \alpha_k^2.
\end{array}
\]
Similarly, we can prove that
\[
\begin{array}{lll}
\sum\limits_{k=1}^{K'} \gamma^k \sum\limits_{r=1}^{k-1} \gamma^{k-r} \alpha_{r-1} \leq \frac{\alpha_0}{1-\gamma}\sum\limits_{k=1}^{K'} \gamma^k,\\[0.2cm]
\sum\limits_{k=1}^{K'} \gamma^k \alpha_{k-1} \leq \frac{1}{2} (\sum\limits_{k=1}^{K'} \gamma^{2k} + \sum\limits_{k=1}^{K'} \alpha_{k-1}^2), \\[0.2cm]
\sum\limits_{k=1}^{K'} \sum\limits_{r=1}^{k-1} \gamma^{k-r} \alpha_{r-1}\alpha_{k-1} \leq \frac{1}{2(1-\gamma)}(\sum\limits_{r=1}^{K'} \alpha_{r-1}^2 + \sum\limits_{r=1}^{K'} \alpha_r^2).
\end{array}
\]
Therefore, $\sum\limits_{k=1}^\infty \|z_i(k) - \overline{z}(k)\|^2 < + \infty$, which means $\lim\limits_{k\rightarrow\infty} \|z_i(k) - \overline{z}(k)\| = 0$ for all $1 \leq i \leq n$. Hence, $\lim\limits_{k\rightarrow\infty} \|x_i(k) - x_j(k)\| = 0$ for all $1 \leq i, j \leq n$.

It follows from (\ref{unbalance1}) and (\ref{algorithm}) that
\[
q_i(k+1) = q_i(k) + [-q_i(k) + \sum\limits_{j\in N_i} a_{ij} [x_j(k) - x_i(k)] + \epsilon y_i(k)]T.
\]
Let $v_i(k) =  [\sum\limits_{j\in N_i} a_{ij} [x_j(k) - x_i(k)] + \epsilon y_i(k)]T$. Then $q_i(k+1) = (1-T)q_i(k) + v_i(k)$. Since $\lim\limits_{k\rightarrow\infty} \|x_j(k) - x_i(k)\| = 0$ and $\lim\limits_{k\rightarrow\infty} y_i(k) = 0$, we have $\lim\limits_{k\rightarrow\infty} v_i(k) = 0$. By iteration, we have $q_i(k+1) = (1-T)^{k+1} q_i(0) + \sum\limits_{s=0}^k (1-T)^{k-s} v_i(s)$. We claim that $\lim\limits_{k\rightarrow\infty} \sum\limits_{s=0}^k (1 - T)^{k-s} v_i(s) = 0$. In fact, since $\lim\limits_{k\rightarrow\infty} v_i(k) = 0$, for any $\epsilon > 0$, there exists $K_1 > T_0$ such that $v_i(s) < \epsilon T$ for all $s \geq K_1$. Then, for $k > K_1$,
\[
\begin{array}{lll}
\sum\limits_{s=0}^k (1-T)^{k-s}v_i(s) = \sum\limits_{s=0}^{K_1} (1-T)^{k-s} v_i(s) + \sum\limits_{s=K_1+1}^k (1-T)^{k-s}v_i(s)\\[0.2cm]
\leq \sum\limits_{s=0}^{K_1} (1-T)^{k-s} v_i(s) + \epsilon T \sum\limits_{s=K_1+1}^k (1-T)^{k-s} \leq \sum\limits_{s=0}^{K_1}(1-T)^{k-s} v_i(s) + \epsilon.
\end{array}
\]
Let $k \rightarrow \infty$, we have $\lim\limits_{k\rightarrow\infty} \sum\limits_{s=0}^{k-s} (1-T)^{k-s} v_i(s) \leq \epsilon$. By the arbitrariness of $\epsilon$, we can conclude that $\lim\limits_{k \rightarrow \infty} \sum\limits_{s=0}^k (1-T)^{k-s} v_i(s) = 0$. Hence, $\lim\limits_{k\rightarrow\infty} q_i(k) = 0$ for all $1 \leq i \leq n$. Thus, $\lim\limits_{k\rightarrow\infty} \|r_i(k) - \overline{z}(k)\| \leq \lim\limits_{k\rightarrow\infty} [\|x_i(k) - \overline{z}(k)\| + \|q_i(k)\|] = 0$. Therefore, $\lim\limits_{k\rightarrow\infty} \|r_i(k) - r_j(k)\| = 0$ and $\lim\limits_{k\rightarrow\infty} q_i(k) = 0$ for all $1 \leq i, j \leq n$, which suggests that there exists a fixed point $x^* \in \mathbf{R}^s$, such that $\lim\limits_{k\rightarrow\infty} \|r_i(k) - x^*\| = \lim\limits_{k\rightarrow\infty} \|x_i(k) - x^*\| = \lim\limits_{k\rightarrow\infty} \|\overline{z}(k) - x^*\| = 0$ for all $1 \leq i \leq n$.

Next, let us prove $x^* \in X$. Define $V(k) = \|\overline{z}(k) - P_X(\overline{z}(k))\|^2$, $k \geq T_0$, where $P_X(\overline{z}(k))$ represents the projection of $\overline{z}(k)$ on $X$. It is clear that
\[
\begin{array}{lll}
V(k+1) &=& \|\overline{z}(k+1) - P_X(\overline{z}(k+1))\|^2\\[0.2cm]
&=& \|\overline{z}(k+1) - P_X(\overline{z}(k)) + P_X(\overline{z}(k)) - P_X(\overline{z}(k+1))\|^2\\[0.2cm]
&\leq& \|\overline{z}(k+1) - P_X(\overline{z}(k))\|^2 + \|\overline{z}(k) - \overline{z}(k+1)\|^2\\[0.2cm]
&=& \|\overline{z}(k+1) - \overline{z}(k) + \overline{z}(k) - P_X(\overline{z}(k))\|^2 + \|\overline{z}(k) - \overline{z}(k+1)\|^2\\[0.2cm]
&=& 2\|\overline{z}(k+1) - \overline{z}(k)\|^2 +  \|\overline{z}(k) - P_X(\overline{z}(k))\|^2\\[0.2cm]
 \ \ \ \ &&+ 2[\overline{z}(k) - P_X(\overline{z}(k))]^T[\overline{z}(k+1) - \overline{z}(k)].
\end{array}
\]
Note that $\overline{z}(k+1) = \overline{z}(k) - \frac{1}{n} \alpha_k T \sum\limits_{i=1}^{2n} [\nabla F(k)]_i =  \overline{z}(k) - \frac{1}{n} \alpha_k T \sum\limits_{i=1}^{n} \nabla f_i(w_i(k))$, we have $\|\overline{z}(k+1) - \overline{z}(k)\|^2 \leq D^2 T^2 \alpha_k^2$. Since $\lim_{k\rightarrow\infty} \overline{z} (k) = x^*$, there must exist $\mu > 0$, such that $\|\overline{z}(k) - P_X(\overline{z}(k))\| \leq \mu$ for all $k$. Hence,

\[
\begin{array}{lll}
&&V(k+1) \\[0.2cm]
&\leq& V(k) + 2 D^2 T^2 \alpha_k^2 - 2[\overline{z}(k) - P_X(\overline{z}(k))]^T \frac{1}{n} \alpha_k T \sum\limits_{i=1}^{n} f_i(w_i(k))\\[0.2cm]
\end{array}
\]
\[
\begin{array}{lll}
&=&  V(k) + 2 D^2 T^2 \alpha_k^2 - \frac{2}{n} \alpha_k T \sum\limits_{i=1}^n[\overline{z}(k) -w_i(k) + w_i(k)\\[0.2cm]
 && \ \ \ \ \ \ \ \ \ - P_X(\overline{z}(k))]^T  \nabla f_i(w_i(k))\\[0.2cm]
&=& V(k) + 2 D^2 T^2 \alpha_k^2 - \frac{2}{n} \alpha_k T \sum\limits_{i=1}^n[\overline{z}(k) -w_i(k)]^T\nabla f_i(w_i(k))\\[0.2cm]
 &&-\frac{2}{n} \alpha_k T \sum\limits_{i=1}^n[ w_i(k) - P_X(\overline{z}(k))]^T  \nabla f_i(w_i(k))\\[0.2cm]
&\leq&  V(k) + 2 D^2 T^2 \alpha_k^2 - \frac{2}{n} \alpha_k T \sum\limits_{i=1}^n[\overline{z}(k) -w_i(k)]^T\nabla f_i(w_i(k))\\[0.2cm]
 &&-\frac{2}{n} \alpha_k T \sum\limits_{i=1}^n[ f_i(w_i(k)) - f_i(P_X(\overline{z}(k)))]\\[0.2cm]
&=& V(k) + 2 D^2 T^2 \alpha_k^2 - \frac{2}{n} \alpha_k T \sum\limits_{i=1}^n[\overline{z}(k) -w_i(k)]^T\nabla f_i(w_i(k))\\[0.2cm]
 &&-\frac{2}{n} \alpha_k T \sum\limits_{i=1}^n[ f_i(w_i(k)) - f_i(\overline{z}(k)) + f_i(\overline{z}(k))- f_i(P_X(\overline{z}(k)))]\\[0.2cm]
&=& V(k) + 2 D^2 T^2 \alpha_k^2 - \frac{2}{n} \alpha_k T \sum\limits_{i=1}^n[\overline{z}(k) -w_i(k)]^T\nabla f_i(w_i(k)) \\[0.2cm]
&&-\frac{2}{n} \alpha_k T \sum\limits_{i=1}^n[ f_i(w_i(k)) - f_i(\overline{z}(k))] -\frac{2}{n} \alpha_k T \sum\limits_{i=1}^n[ f_i(\overline{z}(k))- f_i(P_X(\overline{z}(k)))]\\[0.2cm]
&=& V(k) + 2 D^2 T^2 \alpha_k^2 - \frac{2}{n} \alpha_k T \sum\limits_{i=1}^n[\overline{z}(k) -w_i(k)]^T\nabla f_i(w_i(k))\\[0.2cm]
 &&-\frac{2}{n} \alpha_k T \sum\limits_{i=1}^n[ f_i(w_i(k)) - f_i(\overline{z}(k))] -\frac{2}{n} \alpha_k T [ f(\overline{z}(k))- f(P_X(\overline{z}(k)))].
\end{array}
\]
Since $\lim\limits_{k\rightarrow\infty} \|\overline{z}(k) - w_i(k)\| = 0$, $\nabla f_i(w_i(k))$ is bounded and $\lim\limits_{k\rightarrow\infty} \alpha_k = 0$, for any $\epsilon_2 > 0$, there exists $T_1 > T_0$ such that
\[
 - \frac{2T}{n} \sum\limits_{i=1}^n[\overline{z}(k) -w_i(k)]^T \nabla f_i(w_i(k)) -\frac{2T}{n} \sum\limits_{i=1}^n[ f_i(w_i(k)) - f_i(\overline{z}(k))] < \epsilon_2 T
\]
and $T \alpha_k < \epsilon_2$ for all $k \geq T_1$. Hence,
\[
\begin{array}{lll}
V(k+1) - V(k) &\leq& 2D^2T\epsilon_2 \alpha_k + T \epsilon_2 \alpha_k - \frac{2\alpha_k T}{n}[f(\overline{z}(k)) - f(P_X(\overline{z}(k)))]\\[0.2cm]
&=& - \frac{2\alpha_k T}{n}[f(\overline{z}(k)) - f(P_X(\overline{z}(k))) - \overline{D} \epsilon_2],
\end{array}
\]
where $\overline{D} = \frac{n}{2} (2D^2 + 1) > 0$. Noted that $\lim\limits_{\|x\|\rightarrow\infty} f(x) = \infty$, there exists $c_1 > 0$ such that $f(x_1) - f(x_2) > \overline{D} \epsilon_2 + 2 \epsilon_3$ for all $x_1 \notin U_1$ and $x_2 \in X$, where $U_1 = \{x | \|x - P_X(x)\| \leq c_1\}$ and $\epsilon_3 > 0$. Take $c_2 = c_1 +2 \epsilon_3$, $U_2 = \{x | \|x - P_X(x)\| \leq c_2\}$. Since $\|\overline{z}(k+1) - \overline{z}(k)\| \leq D^2 T^2 \alpha_k^2$ and $\lim\limits_{k\rightarrow\infty} \alpha_k = 0$, there must exist $T_2 > T_1$ such that $\|\overline{z}(k+1) - \overline{z}(k)\| < \epsilon_3$ for all $k > T_2$.
 When $k > T_2$, if $\overline{z}(k) \in U_1$, i.e., $\|\overline{z}(k) - P_X(\overline{z}(k))\| \leq c_1$, then
 \[
 \begin{array}{lll}
\|\overline{z}(k+1) - P(\overline{z}(k+1))\| \\[0.2cm]
\leq \|\overline{z}(k+1) - \overline{z}(k)\| + \|\overline{z}(k) - P_X(\overline{z}(k))\|
 + \|\overline{z}(k) - \overline{z}(k+1)\|\\[0.2cm]
 \leq 2\epsilon_3 + c_1 = c_2,
 \end{array}
 \]
this is to say that $\overline{z}(k+1) \in U_2$. If $\overline{z}(k) \notin U_1$ but $\overline{z}(k) \in U_2$, then $V(k+1) - V(k) < 0$, which suggests that
$\overline{z}(k+1) \in U_2$. If $\overline{z}(k) \notin U_2$, then $V(k+1) - V(k) < - \frac{4\alpha_k T}{n} \epsilon_3$. Since $\sum\limits_{k=1}^\infty \alpha_k = \infty$,
there must exist $T_3 > T_2$ such that $\overline{z}(k) \in U_2$ for all $k \geq K_3$. Hence, we can conclude that $\|\overline{z}(k) - P_X(\overline{z}(k))\|
\leq c_1 + 2 \epsilon_3$ for all $k \geq K_3$. Letting $\epsilon_3 \rightarrow 0$, we have $\lim\limits_{k\rightarrow\infty} \|\overline{z}(k) - P_X(\overline{z}(k))\| \leq c_1$,
i.e., $\lim\limits_{k\rightarrow\infty} \overline{z}(k) \in U_1$, which yields that
$\lim\limits_{k\rightarrow\infty} [f(\overline{z}(k)) - f(P_X(\overline{z}(k)))] \leq \overline{D}\epsilon_2$.
Letting $\epsilon_2 \rightarrow 0$, we have $\lim\limits_{k\rightarrow\infty} [f(\overline{z}(k)) - f(P_X(\overline{z}(k)))] = 0$. Since $\lim_{k\rightarrow\infty} \|\overline{z}(k) - x^*\| = 0$ and $f_i(x)$ is continuous, we can arrive at $\|f(x^*) - f(P_X(x^*))\| = \lim\limits_{k\rightarrow\infty} [f(\overline{z}(k)) - f(P_X(\overline{z}(k)))] = 0$, i.e., $f(x^*) = f(P_X(x^*))$, which implies $x^* \in X$. The proof of Theorem 1 is finally completed.

\begin{remark}\rm
In \cite{Xi2017}, the distributed first-order multi-agent optimization for unbalanced network was studied. This paper extends this result to the second-order multi-agent networks. In addition, the convergence analysis method in \cite{Xi2017} based on the fact that $\sum\limits_{n=1}^\infty a_n b_n < + \infty$ and $\sum\limits_{n=1}^\infty a_n = \infty$ implies $\lim\limits_{n\rightarrow\infty} b_n = 0$ doesn't work in fact. For example $a_n = \frac{1}{n}$ for all $n$, $b_n = 1$ if $n = k^2$ for some positive integer $k$, otherwise, $b_n = 0$. Clearly, $\sum\limits_{n=1}^\infty a_n b_n < + \infty$ and $\sum\limits_{n=1}^\infty a_n = \infty$, but  $\lim\limits_{n\rightarrow\infty} b_n$ doesn't exist. In this paper, we introduce an alternative method to analyze the convergence of the second-order multi-agent optimization algorithm.
\end{remark}

\section{Conclusions}

The distributed second-order multi-agent optimization problem was considered in this paper, where the underlying networks were assumed to be directed and unbalanced. A new distributed algorithm with switching mechanism was introduced to guarantee the boundedness of the system states. The closed-loop system was divided into two subsystems by a transformation, one of which is independent on another. This leads to the convenience of analysis. We then proved that if the communication network is strongly connected, the position states of all agents can converge to the optimal solution of the team objective function without the assumption of the boundedness of the gradient of private function.\\ 

{\bf References}





\end{document}